\documentclass[]{spie}  

\usepackage{amsmath,amsfonts,amssymb}
\usepackage{graphicx}
\usepackage[colorlinks=true, allcolors=blue]{hyperref}

\title{Forward modelling coronagraphic images with a fully physical, differentiable digital twin of MagAO-X: first laboratory results}

\author[a]{Matthijs Mars}
\author[a,b]{Sebastiaan Y. Haffert}
\author[a]{Louis Desdoigts}
\author[c]{Joseph D. Long}
\author[a]{Rico Landman}
\author[b]{Jared R. Males}
\author[b]{Laird M. Close}
\author[b]{Kyle Van Gorkom}
\author[b,d,e,f]{Olivier Guyon}
\author[g]{Alexander D. Hedglen}
\author[b,e]{Sandrine Juillard}
\author[d]{Jennifer Lumbres}
\author[h]{Lauren Schatz}

\affil[a]{Leiden Observatory, Leiden University, PO Box 9513, 2300 RA Leiden, The Netherlands}
\affil[b]{Steward Observatory, The University of Arizona, 933 North Cherry Avenue, Tucson, AZ, USA}
\affil[c]{Center for Computational Astrophysics, Flatiron Institute, 162 5th Avenue, New York, NY, USA}
\affil[d]{Wyant College of Optical Sciences, The University of Arizona, 1630 E University Blvd, Tucson, AZ, USA}
\affil[e]{Subaru Telescope, National Astronomical Observatory of Japan, 650 N. A'ohoku Place, Hilo, HI, USA}
\affil[f]{Astrobiology Center, National Institutes of Natural Sciences, 2-21-1 Osawa, Mitaka, Tokyo, Japan}
\affil[g]{Northrop Grumman Corporation, 600 South Hicks Road, Rolling Meadows, IL, USA}
\affil[h]{Starfire Optical Range, Kirtland Air Force Base, Albuquerque, New Mexico, USA}
\authorinfo{Further author information: (Send correspondence to M.M.)\\M.M.: E-mail: academic@matthijsmars.com}

\begin{document}
\maketitle

\begin{abstract}
Post-processing of high contrast imaging data relies on an accurate model of the stellar point spread function (PSF).
Current techniques build this model from the science images themselves, using observational diversity (e.g., angular, spectral or polarimetric diversity), which can cause self-subtraction of the companion signal and constrains the observing strategy.
Telemetry-based forward modelling instead builds the stellar PSF model from wavefront sensor data that is already recorded during the observation.
The wavefront sensor measures the coherent starlight and can therefore be used to create a PSF model that only models the stellar light and does not reproduce the incoherent light of a companion.

We present a fully physical and differentiable digital twin of the focal plane low-order wavefront sensor (FLOWFS) and the coronagraphic science beam of the MagAO-X instrument, implemented in \texttt{dLux}, and calibrate it on laboratory data.
When fitted directly to the science images, the model reproduces the coronagraphic PSF down to the photon and read noise floor of the data.
When instead forward modelled from the FLOWFS telemetry alone, the residuals reach $6\times10^{-5}$ of the stellar peak at $5\ \lambda/D$, a factor of 5 below the raw contrast, with the remaining residual set by how well the wavefront estimate transfers from the FLOWFS branch to the science branch of the model.
An injected companion at $5\ \lambda/D$ with a peak contrast of $10^{-3}$ is recovered without measurable self-subtraction.
We discuss the model improvements currently under development and the path towards on-sky validation.
\end{abstract}


\section{INTRODUCTION}
\label{sec:intro}
High contrast imaging aims to detect faint circumstellar objects at small angular separations from their host star, whose point spread function (PSF) is orders of magnitude brighter and buries the object's signal.
To overcome this, high contrast imagers combine coronagraphs, which block as much of the on-axis stellar light as possible \cite{kenworthyHighContrastCoronagraphy2025}, wavefront control, which removes the aberrations introduced by atmospheric turbulence and optimises the wavefront for coronagraphic rejection \cite{guyonExtremeAdaptiveOptics2018}, and post-processing, which attempts to distinguish between the bright stellar light and the faint light of the object of interest.

Despite extreme adaptive optics (XAO) systems correcting for most of the aberrations introduced by atmospheric turbulence, residual wavefront errors persist after correction, and further aberrations are introduced by vibrations of the instrument and telescope as well as slowly evolving non-common path aberrations. These degrade the coronagraphic performance and cause leakage of stellar light into the science images \cite{guyonExtremeAdaptiveOptics2018}.
Current post-processing techniques often model this stellar leakage, requiring specific observing strategies to build a model of the stellar PSF that does not contain the companion signal, which would otherwise self-subtract.
These strategies rely on angular, spectral, polarimetric, or coherence diversity, or on observing reference stars, constraining the instrument setup and the observing strategy.

Telemetry-based forward modelling uses wavefront-sensor data recorded during the observation to reconstruct residual aberrations and propagate them through an instrument model to predict the stellar PSF \cite{guyonHighContrastImaging2022}.
Because the prediction is driven by wavefront telemetry rather than by the science images themselves, the stellar model is constrained to coherent starlight and does not absorb incoherent companion flux.
As a result, no angular/spectral/polarimetric diversity or reference star is required, although such diversity can still be combined with this approach.
The challenge then shifts to model fidelity: the subtraction quality is set by how accurately the instrument response is represented.
In previous work, we demonstrated the use of a digital twin for low-order wavefront sensing and control using on-sky MagAO-X experiments \cite{marsFLOWFS2026}.
Fully differentiable models, or digital twins, of the instrument have also been used successfully for PSF modelling on space-based instruments \cite{desdoigtsAMIGODataDrivenCalibration2025,fengExoplanetDetectionDifferentiable2025}.

In this work, we present ongoing work on a fully physical and differentiable digital twin of the MagAO-X instrument \cite{malesMagAOXCommissioningResults2024} that models both the focal plane low-order wavefront sensor (FLOWFS) and the coronagraphic science beam of the instrument.
First, the digital twin is calibrated using dedicated calibration data including both images from the science and FLOWFS camera.
For our science observations, we reconstruct the wavefront from the FLOWFS images alone, forward model it through the science branch, and subtract the resulting PSF from the science images.
We demonstrate the ability of our fully physical and differentiable digital twin to reproduce the stellar PSF using lab data and show that we can use forward modelling to retrieve a synthetic injected companion without oversubtraction. 

In \autoref{sec:dt}, we describe the layout of MagAO-X, the digital twin and its parameterisation, the fitting procedure, and the forward-modelling strategy.
In \autoref{sec:calibration}, we calibrate the model on laboratory data, first with a flat wavefront and then with injected low-order Zernike aberrations.
In \autoref{sec:fm}, we forward model the science PSF from FLOWFS telemetry and analyse the residuals, and in \autoref{sec:companion} we inject a companion and measure how well it is recovered.
We conclude in \autoref{sec:discussion} by discussing what limits the current model and how we plan to take this technique on-sky.

\section{A digital twin of MagAO-X}
\label{sec:dt}

\subsection{Instrument Layout}

MagAO-X is an extreme adaptive optics high contrast imager on the 6.5\,m Magellan Clay telescope at Las Campanas Observatory \cite{malesMagAOXCommissioningResults2024}.
The primary AO correction is done by a woofer-tweeter pair consisting of an ALPAO-97 deformable mirror (DM) and a Boston Micromachines 2K DM \cite{malesMagAOXProjectStatus2018,closeOpticalMechanicalDesign2018,malesMagAOXCurrentStatus2022,malesMagAOXCommissioningResults2024} driven by a pyramid wavefront sensor. 
In the science arm, we have a Boston Micromachines 1K functioning as a dedicated non-common path correction (NCPC) DM upstream of the coronagraph optics. 
In this work, we use a Lyot-style reflective coronagraph (a chrome dot with a diameter of $252\,\mu$m, $\sim 2.19 \lambda / D$ at 900\,nm) that not only blocks the on-axis stellar light but redirects it to a dedicated FLOWFS camera that is used for low-order wavefront sensing. 
Since this WFS receives rejected light, it has a very large signal and can therefore run at kHz speeds. 
The FLOWFS camera is deliberately defocused, which breaks degeneracies in the response to the different aberration modes.
A schematic diagram of the science arm of MagAO-X is shown in \autoref{fig:diagram} together with the model description.

\subsection{The Model}
\label{sec:model}
\begin{figure}[h!]
	\centering
	\includegraphics[width=0.92\textwidth]{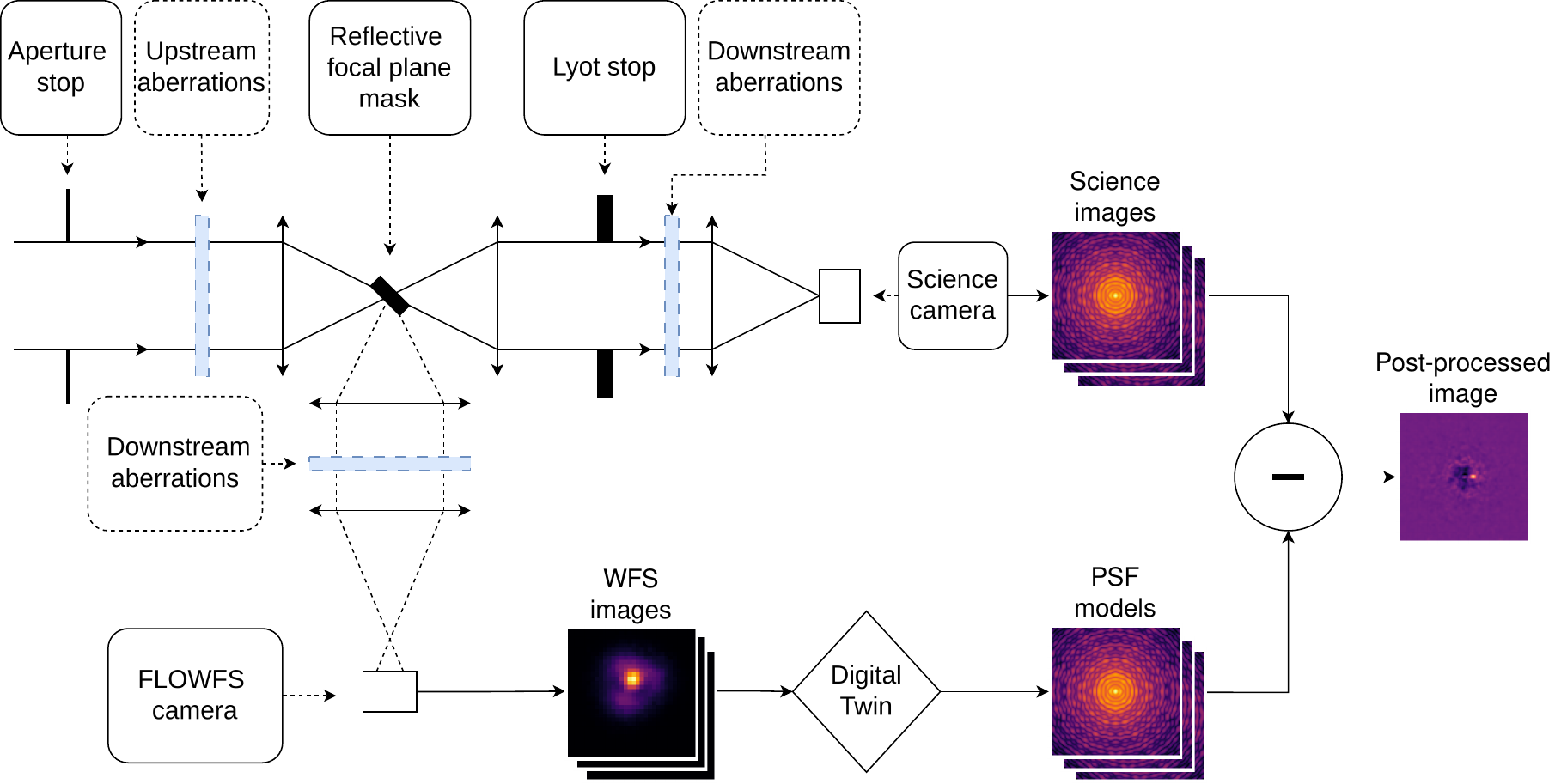}
	\caption[diagram]{
	The optical layout that the digital twin reproduces, and the post-processing pipeline it enables.
	The beam passes through the aperture stop and is brought to a focus on the reflective focal plane mask.
	On-axis stellar light rejected by the mask is reflected to the intentionally defocused FLOWFS camera, while the remaining light continues through the Lyot stop to the science camera.
	The model parameterises aberration planes upstream and downstream of the focal plane mask, and downstream of the reflection into the FLOWFS arm.
	Wavefront aberrations are reconstructed from the FLOWFS images and forward modelled using the digital twin through the science branch to produce a PSF model for every science exposure.
	Subtracting these models from the science images removes the stellar light.
	Since the model only produces coherent stellar light and its aberrations are inferred from the wavefront sensor telemetry, the PSF model is not sensitive to the incoherent signal of a companion.
	}
	\label{fig:diagram}
\end{figure}

Our digital twin of the MagAO-X science arm is fully physical and differentiable and is implemented in \texttt{dLux}\cite{desdoigtsDifferentiableOpticsDLux2023}, a physical optics framework built in JAX. 
Every operation in the optical propagation is differentiable, so gradients of any output with respect to any model parameter are available by automatic differentiation \cite{baydinAutomaticDifferentiationMachine2018}, and do not need to be derived analytically as in our previous work \cite{marsFLOWFS2026}.
The model shares a single set of upstream layers between the two cameras and then branches, so that the two propagations differ only downstream of the focal plane mask.
A schematic visualisation of the model can be found in \autoref{fig:diagram} and it consists of the following elements:
\begin{enumerate}
	\item The aperture stop, with fitted translation, rotation, scale and shear. 
	\item Upstream phase and amplitude aberrations.
	\item The DM, which models the injected Zernike aberrations as well as residual low-order aberrations.
	\item Propagation onto the reflective focal plane mask.
	\item Either the Lyot stop and the science branch, or the reflection to the FLOWFS branch.
	\item Downstream aberrations, separately for each camera.
	\item Imaging on the detector, including jitter and a constant background.
\end{enumerate}

To describe the instrument response, we split the model parameters into static and dynamic terms.
The static terms capture properties that are approximately constant over an observing sequence: alignment and distortion of the aperture stop, focal plane mask and Lyot stop, focal lengths, and high-order aberrations in the upstream and downstream parts of the optical system.

The dynamic terms capture frame-to-frame changes from residual low-order aberrations (e.g. bench seeing, vibrations and quasi-static NCPA), plus detector normalisation.
In this work, we model these changes with 9 Zernike coefficients on the NCPC DM, a per-camera downstream tip/tilt, and per-frame flux and background.
For now, we only fit varying low-order modes: modelling time-varying high-order aberrations would require a dedicated pyramid-WFS model, since FLOWFS is not sensitive to them.
The reflective focal plane mask is small, and acts as a spatial filter that limits the amount of spatial information that the FLOWFS can sense.
We therefore restrict the aberrations applied by the DM to the first 9 Zernike modes, excluding piston (Noll indices 2 to 10).
A summary of all parameters is given in \autoref{tab:params}.

\begin{table}[ht]
	\caption{The parameterisation of the digital twin; the static parameters describe the instrument, the per-frame parameters describe the state of the wavefront and the detector during a single exposure.}
	\label{tab:params}
	\begin{center}
	\begin{tabular}{|l|p{9.5cm}|}
	\hline
	\rule[-1ex]{0pt}{3.5ex}  \textbf{Element} & \textbf{Parameters}  \\
	\hline
	\multicolumn{2}{|l|}{\rule[-1ex]{0pt}{3.5ex}\textit{Static, shared by both cameras}} \\
	\hline
	\rule[-1ex]{0pt}{3.5ex}  Aperture stop & Translation, rotation, scale, shear   \\
	\rule[-1ex]{0pt}{3.5ex}  Focal plane mask & Rotation, scale, shear   \\
	\rule[-1ex]{0pt}{3.5ex}  Lyot stop & Translation, rotation, scale, shear   \\
	\rule[-1ex]{0pt}{3.5ex}  Upstream aberrations & 1600 Fourier modes in amplitude and phase  \\
	\hline
	\multicolumn{2}{|l|}{\rule[-1ex]{0pt}{3.5ex}\textit{Static, per camera}} \\
	\hline
	\rule[-1ex]{0pt}{3.5ex}  Downstream aberrations & 1600 Fourier modes in amplitude and phase (science camera); 9 Zernike modes (FLOWFS) \\
	\rule[-1ex]{0pt}{3.5ex}  Imaging & focal length, jitter  \\
	\hline
	\multicolumn{2}{|l|}{\rule[-1ex]{0pt}{3.5ex}\textit{Dynamic, per frame}} \\
	\hline
	\rule[-1ex]{0pt}{3.5ex}  Low-order aberrations & 9 Zernike modes (Noll 2 to 10)  \\
	\rule[-1ex]{0pt}{3.5ex}  Downstream aberrations & Tip and tilt  \\
	\rule[-1ex]{0pt}{3.5ex}  Normalisation & Flux, background  \\
	\hline
	\end{tabular}
	\end{center}
\end{table}

\subsection{Fitting the Model}
\label{sec:fitting}

We fit the model by minimising the mean squared z-score of the residuals, that is, the residuals normalised by their expected noise:
\begin{equation}
\label{eq:loss}
\mathcal{L}(\boldsymbol{\theta}) = \frac{1}{N} \sum_i
\left( \frac{y_{\text{obs},i} - y_{\theta,i}}{\sigma_i(\boldsymbol{\theta})} \right)^2 ,
\qquad
\sigma_i(\boldsymbol{\theta}) = \sqrt{ \sigma_{\text{photon}, \theta, i}^2 + \sigma_{\text{read}}^2 } ,
\qquad
\sigma_{\text{photon}, \theta, i} \approx \sqrt{y_{\theta,i} }
\end{equation}
with $y_{\text{obs},i}$ and $y_{\theta,i}$ the $i$-th pixel of the observed and forward modelled images, $\sigma_{\text{read}}$ the read noise of the relevant camera, $\sigma_{\text{photon}, \theta, i}$ the photon noise as approximated by the predicted PSF model, and $N$ the number of pixels summed over all exposures.
The noise estimate therefore contains both the photon noise of the model and the read noise of the detector, so that bright pixels near the core are automatically down-weighted relative to the faint pixels in the halo that carry the contrast information.
Saturated pixels and their immediate neighbours are masked.

The loss is minimised through gradient descent, using per-parameter learning rates.
Parameters that are poorly constrained early in the optimisation, such as the amplitude aberrations, are held fixed and then ramped in after a delay, which we found to converge more reliably than optimising everything at once.
With the optimisation we can fit the shared static parameters jointly over several observations while also optimising the per-frame dynamic parameters individually.

\subsection{Forward-Modelling Strategy}
\label{sec:strategy}

The process of forward modelling the coronagraphic images using our digital twin has three steps:
\begin{enumerate}
	\item Calibrate the static model on laboratory data, fitting both cameras jointly so that a single set of static parameters explains them both.
	\item Reconstruct the wavefront from the FLOWFS images alone using the calibrated model, recovering the 9 Zernike coefficients for every frame.
	\item Freeze those coefficients, propagate them through the science branch of the model, and fit only the per-frame tip/tilt, flux and background on the science camera.
\end{enumerate}
The resulting PSF model is subtracted from the science image.

The third step is the one that matters for post-processing.
The wavefront that generates the science PSF model comes entirely from the FLOWFS telemetry, and the only quantities that the science images are allowed to constrain are three per-frame nuisance parameters (a tip/tilt of the final image, a flux normalisation, and a constant background) and a static low-order phase offset shared across all frames (\autoref{sec:fm}).
Using this strategy, the model is set up to only produce coherent stellar light, and because the aberrations are inferred from the wavefront sensor rather than the science images, the fit is not sensitive to incoherent light such as a companion's signal.

\section{CALIBRATION}
\label{sec:calibration}

We calibrate the model on laboratory data taken with the MagAO-X internal source using a calibration dataset consisting of a sweep through the 9 Zernike modes we want to sense and control.
The science camera and the FLOWFS camera are read out simultaneously for every DM state at 20\,Hz and 4\,kHz respectively, using a narrowband 875\,nm filter for the science camera and a broadband filter with a central wavelength of 908\,nm for the FLOWFS camera.
The camera framerate and camera gain are chosen to optimise the signal-to-noise ratio as well as to run as fast as possible in order to try and freeze the aberrations on a frame-by-frame basis. 
We model both branches of the optical system monochromatically at the central wavelength of each respective filter.
Science frames are dark-subtracted, corrected for detector gain, and cropped to $150\times150$ pixels.
All science camera residuals in the next sections are normalised using the stellar peak flux of non-coronagraphic lab data (data without the focal plane coronagraph) such that they are an indication of the contrast between the residual features and the stellar intensity.
The FLOWFS images are normalised by the peak flux of the reflected on-axis stellar light when no additional aberrations are introduced. 
To compare the different results we look at the standard deviation of the residuals in annuli around the PSF centre, normalised by the stellar peak flux. 
This gives us an estimate of the peak flux necessary for a 1-$\sigma$ detection.
We also compare our residuals to the raw intensity in the images as well as the noise floor as set by the photon and read noise. 

\subsection{Modelling bench seeing}
\label{sec:flat}

\begin{figure}[ht]
	\centering
	\includegraphics[width=0.45\textwidth]{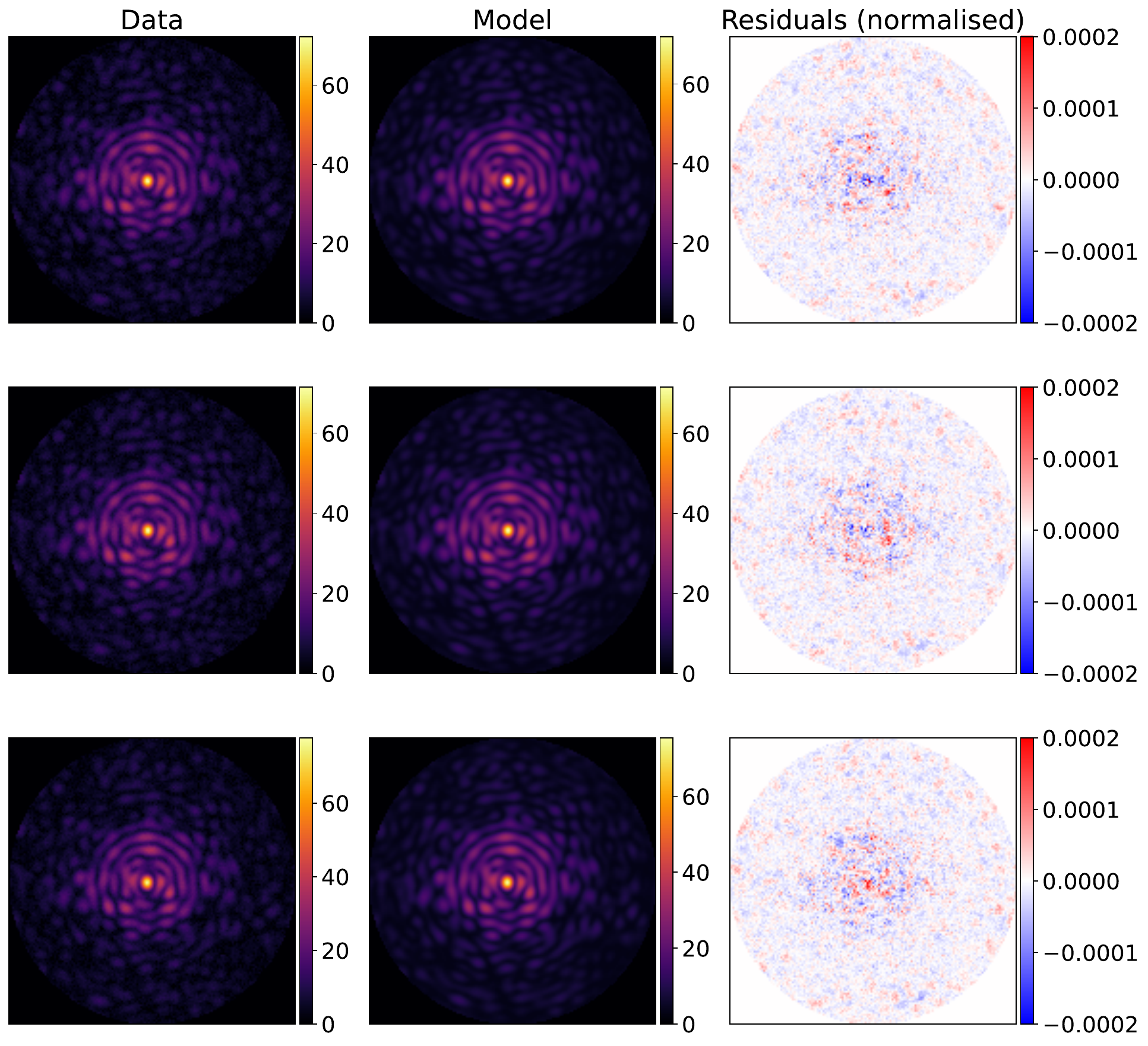} 
	\includegraphics[width=0.40\textwidth]{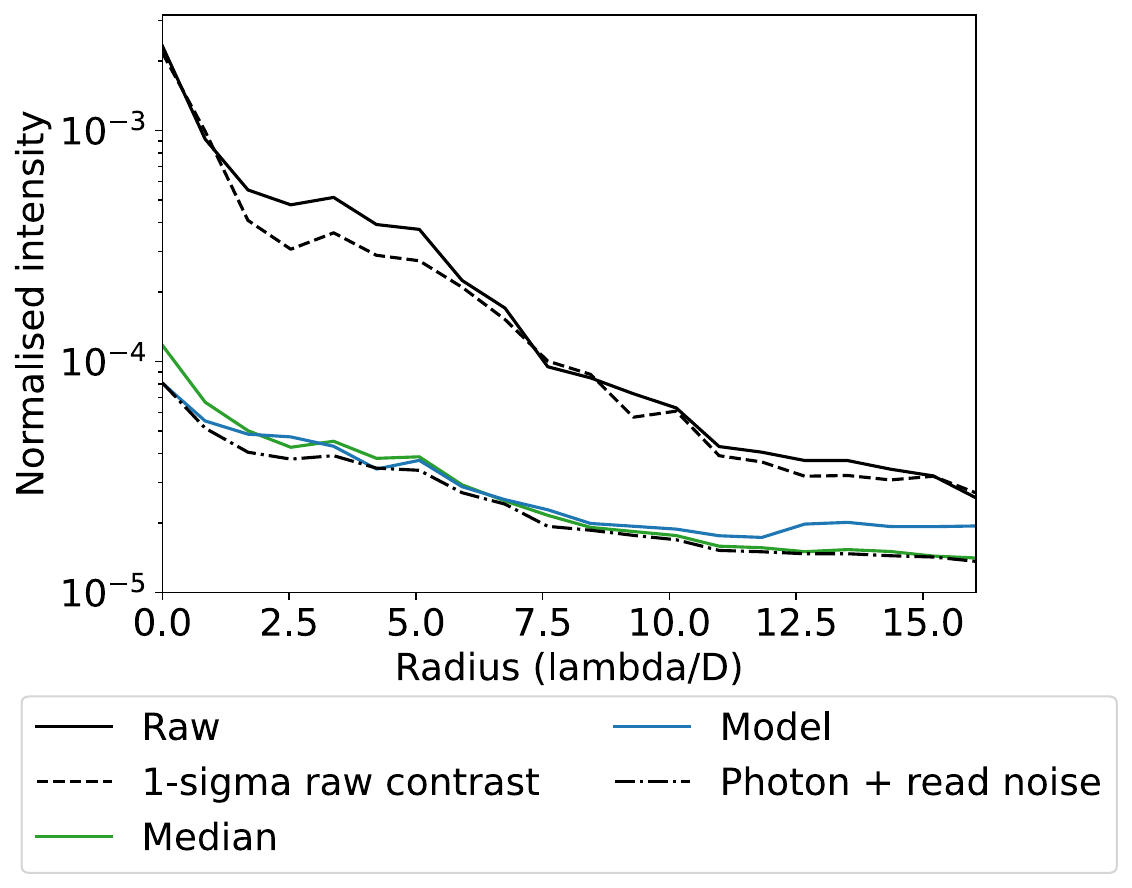}
	\caption{\label{fig:flat}
		The performance of our digital twin on coronagraphic lab data where the only variation between frames is bench seeing.
		Left: three exposures of the coronagraphic science PSFs (data), the fitted model, and the residuals normalised by the stellar peak (non-coronagraphic).
		The residuals are mostly uncorrelated and dominated by noise.
		Right: the intensity of the residual per annulus, normalised by the peak of the unocculted stellar PSF.
		We show the raw normalised intensity of the coronagraphic PSF as well as the standard deviation of the residuals with no subtraction, subtraction using our model PSF, and subtraction using a median over 8 exposures (not including the true exposure). 
		The performance limit of the modelling approaches is set by the photon and read noise in the observation and is calculated from the true frame. 
		Both the fitted model and a median subtraction accurately reproduce the features of the coronagraphic PSF approaching the limit set by the noise floor.
	}
\end{figure}

We first fit the model to exposures taken with no aberration injected on the DM to demonstrate that our model can capture both the static behaviour of our instrument as well as small variations introduced through bench seeing. 
\autoref{fig:flat} shows three such exposures each separated by approximately one second, the fitted model, and the normalised residuals as well as the standard deviation of the residuals as a function of radius.
Our model reproduces the coronagraphic PSFs and variations due to bench seeing accurately with the residuals looking mostly unstructured and noise dominated.
When looking at the normalised intensity, we see that the standard deviation of the residuals approaches the estimated performance limit set by the photon and read noise of the observation at $4 \times 10^{-5}$ at $5 \lambda/D$ compared to $3 \times 10^{-4}$ for the raw 1-$\sigma$ contrast. 
Subtracting using a median of the science frames performs very similarly, an indication of the stability of the instrument, though this modelling strategy is very prone to oversubtraction and is therefore sub-optimal. 

\subsection{Modelling low-order aberrations}
\label{sec:joint}

For our model to be used for forward modelling, it needs to be able to reproduce the effects of the low-order aberrations that we are trying to correct for and do so consistently between the science and FLOWFS camera. 
Therefore we calibrate a model that jointly fits the science camera and FLOWFS camera responses to three different DM states: one without added aberrations, one with additional defocus, and one with additional astigmatism. 
These three frames were chosen to keep the optimisation from taking up too much time while showing robustness to a range of aberrations, though in our experiments adding more exposures with different aberrations to the optimisation did not seem to harm the performance significantly other than increasing the computation time.

\begin{figure}[ht]
	\centering
	\includegraphics[height=0.45\textwidth]{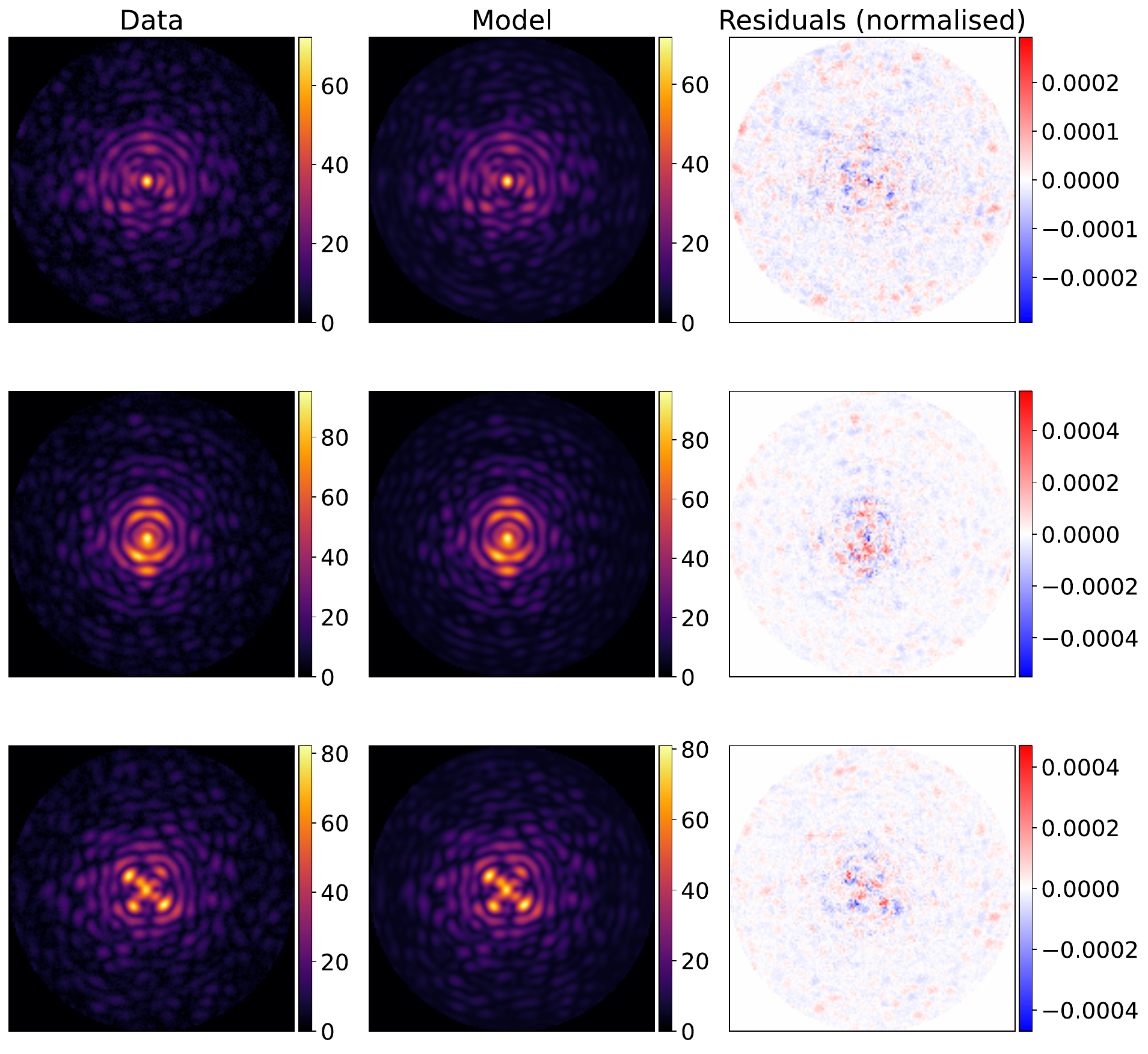} 
	\includegraphics[height=0.45\textwidth]{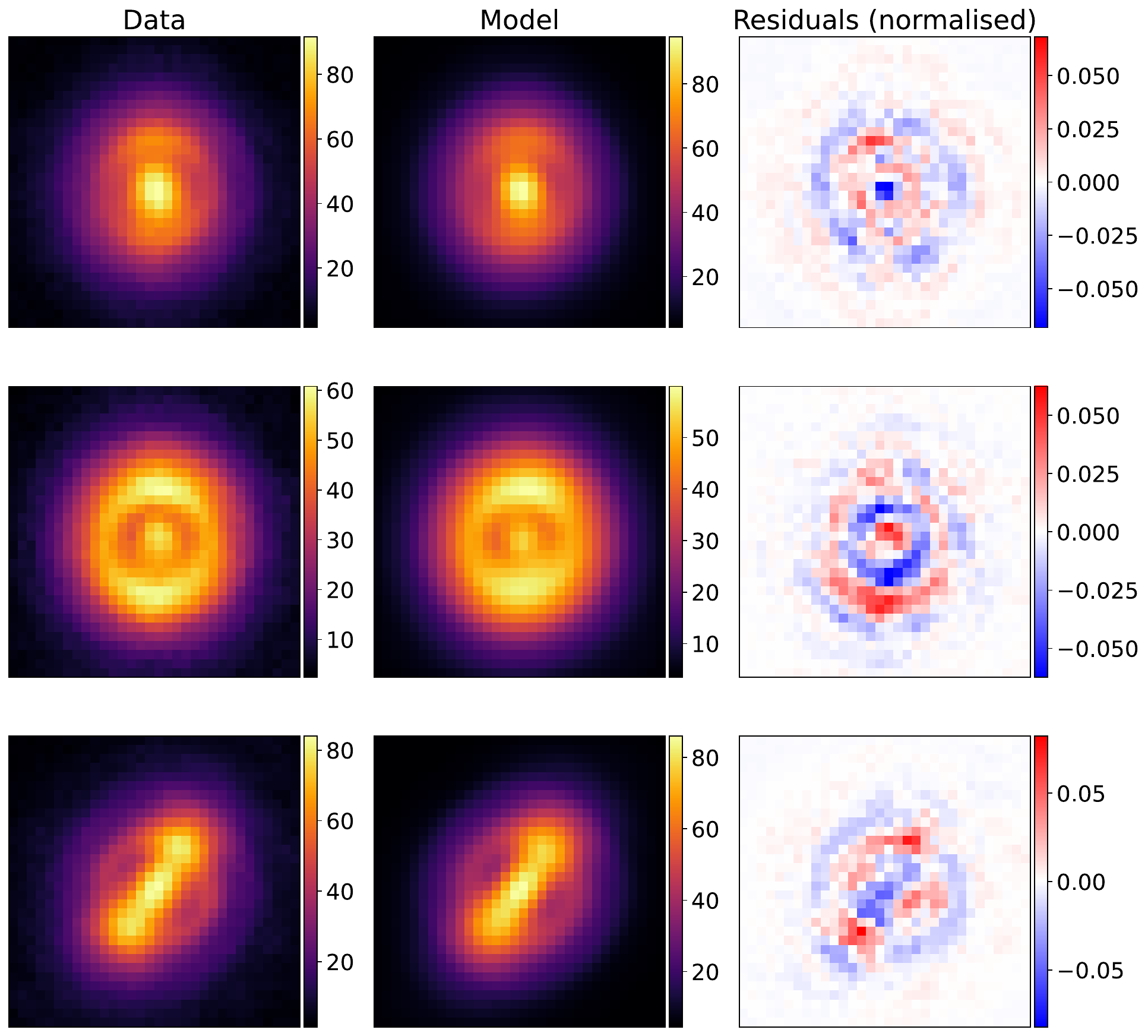} 
	\caption[joint]{\label{fig:joint}
		Left: the science camera images for three different DM states, showing data, model and normalised residuals.
		Right: the FLOWFS camera images for the same three states.
		One set of static parameters reproduces both cameras at once as the injected aberration changes. 
		The residuals remain mostly unstructured and noise dominated in the science camera and the slight increase in residuals is predominantly due to the leakage introduced with these aberrations increasing the amount of photon noise in the images.
		In the FLOWFS residuals we see some structure that seems to be circular and common in all three frames at roughly 5\% of the peak stellar flux. 
		The corresponding residual per annulus is plotted in \autoref{fig:fm} as ``Model (w/ aberrations)''.
	}
\end{figure}

\autoref{fig:joint} shows the reconstructions on both cameras for the three representative DM states.
For the science frame with no aberrations introduced we see almost no difference compared to when we just modelled the bench seeing. 
The coronagraphic PSFs with additional aberrations that cause leakage of stellar light through the coronagraph are also modelled accurately with only slightly higher residuals, which is in part due to the fact that the leaked stellar light also comes with an increased amount of photon noise. 
While the morphology of the FLOWFS PSFs is reproduced reasonably well, there seems to be a common circular residual structure in each of these frames at about 5\% of the peak flux.
In our experiments this effect persisted when modelling the FLOWFS images using broadband light (as we are using a broadband filter). 
We believe that this might be due to a reflection of off-axis light on the substrate that seats the reflective focal plane mask and this is part of ongoing investigations.

The normalised intensity of the residual of the non-aberrated frame is plotted in \autoref{fig:fm}, alongside the forward-modelling result that they should be compared against.
Note that this is fit directly to the science images, so it is not a post-processing result, but it sets the upper bound on what the forward model of \autoref{sec:fm} could achieve: it tells us how well the model can possibly represent these data when it is allowed to look at them.
When comparing the residuals to the model calibrated using just bench seeing, we see that there is a slight increase in the residuals around $3\lambda /D $ as well as out to the outer part of the PSF. 
This could indicate a minor mismatch between how our digital twin models the DM and thus the low-order aberrations and this is another of the things we are working on. 

\section{FORWARD MODELLING THE SCIENCE PSF}
\label{sec:fm}

\begin{figure}[ht]
	\centering
	\includegraphics[width=0.45\textwidth]{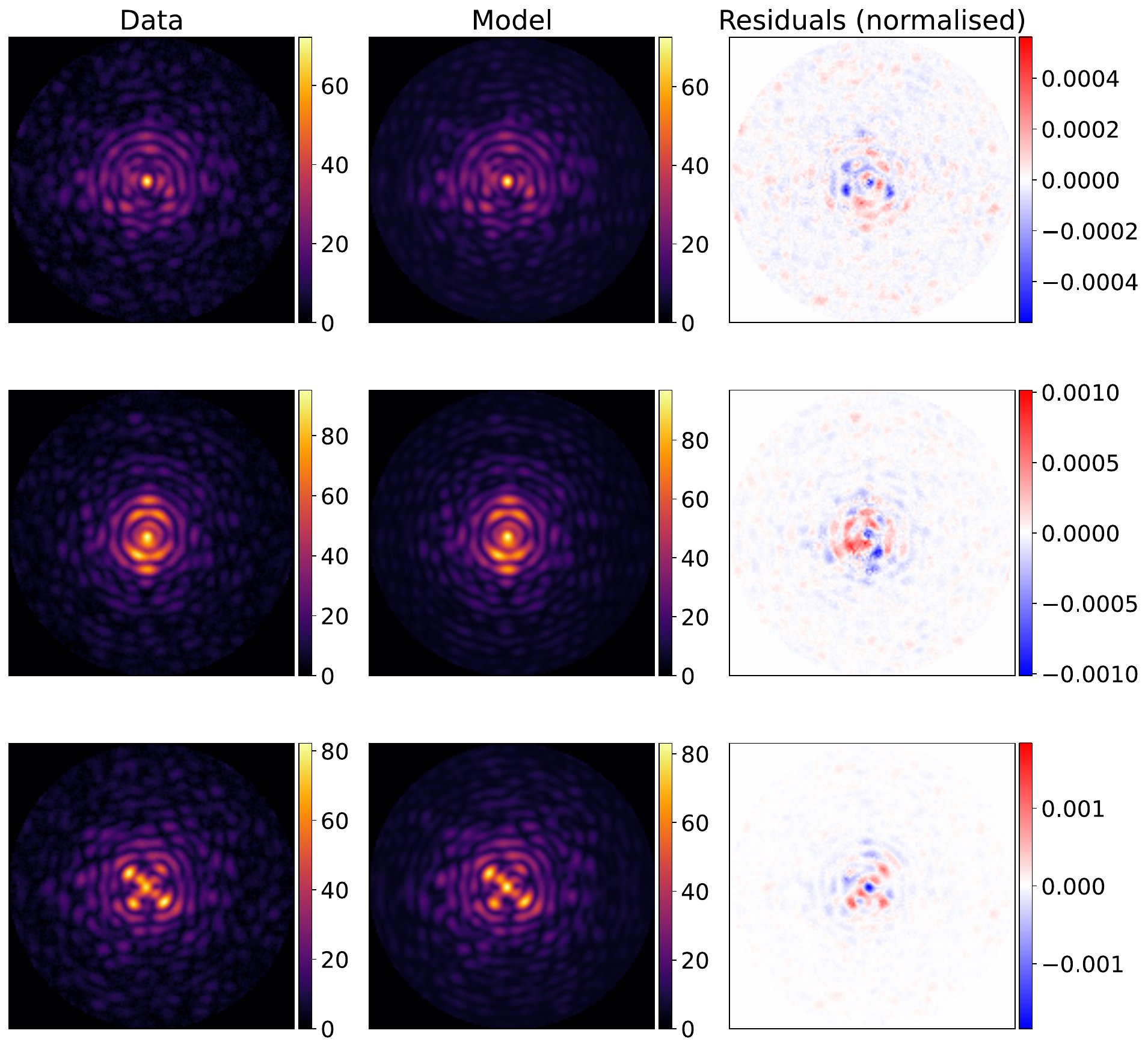} 
	\includegraphics[width=0.40\textwidth]{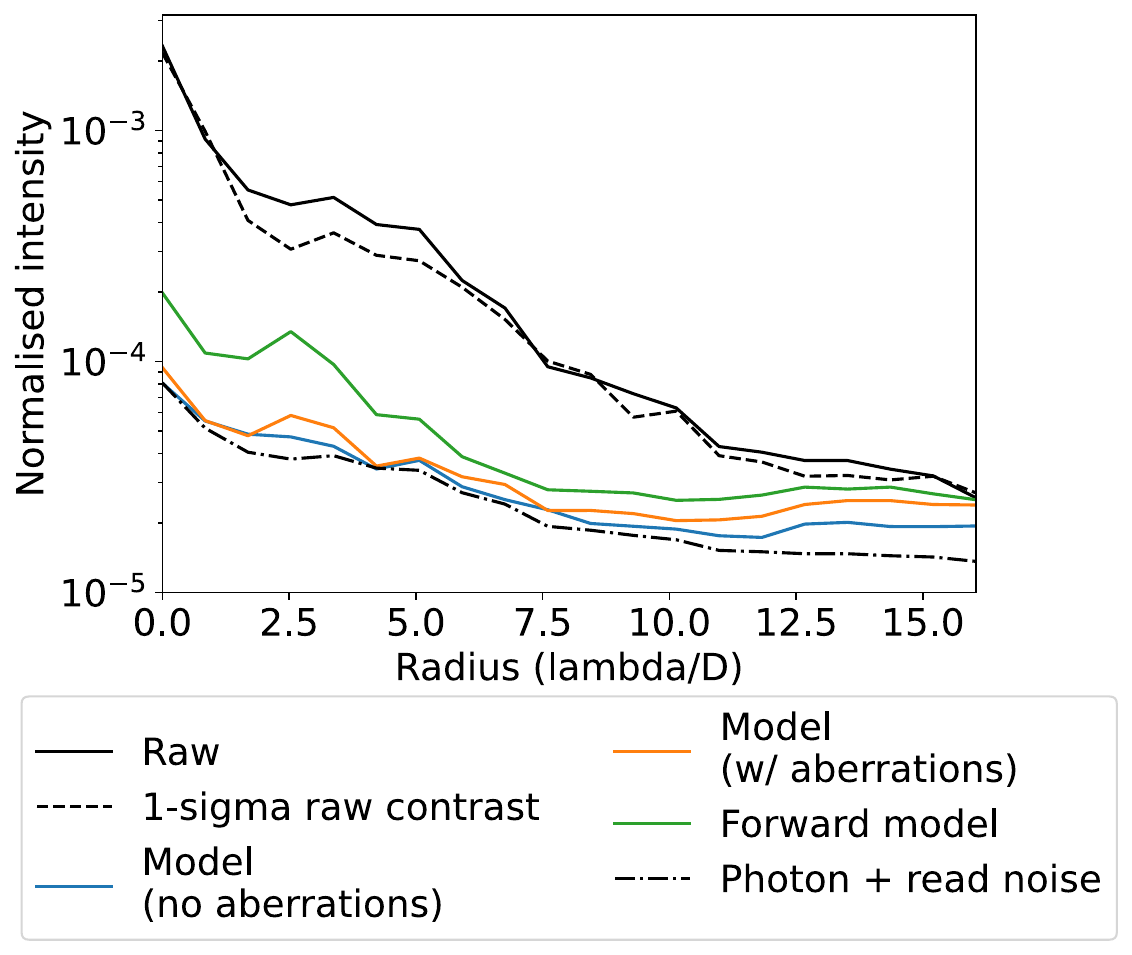}
	\caption[fm]{\label{fig:fm}
		Forward modelling the science PSF from FLOWFS telemetry.
		Left: three science exposures, the PSF predicted from the FLOWFS reconstruction, and the normalised residuals.
		Unlike \autoref{fig:joint}, the residuals show clear structure in the core, which is the signature of model mismatch rather than noise.
		Right: the residual per annulus for the forward model, compared with the direct fits to the science camera with and without additional aberrations and with the raw contrast.
		The forward model gains a factor of 5 in normalised intensity at $5\ \lambda/D$ compared to the 1-$\sigma$ raw contrast, but does not reach the same performance as fitting the low-order aberrations directly to the science data.
	}
\end{figure}

Using the calibrated static model, we reconstruct the coefficients of the 9 Zernike modes that model the low-order aberrations using only the FLOWFS images for the 99 frames of the aberration sweep.
Since we have multiple FLOWFS frames per science frame, we take 10 FLOWFS frames throughout the science exposure and average the reconstructed coefficients over them to attempt to reduce the variation due to bench seeing.
These coefficients are then frozen and propagated through the science branch of the calibrated model, with only the per-frame tip/tilt, flux and background left free on the science camera, as described in \autoref{sec:strategy}.
Since there seemed to be a static discrepancy between the coefficients recovered from the FLOWFS images and the science images, we additionally fit a low-order phase offset (9 Zernike modes) that is shared across all frames.
Because this offset is static, fitted jointly across the 99 frames with different DM states, and restricted to low-order modes, it should not be able to reconstruct incoherent speckles or cause oversubtraction.
We note that these frames come from the same dataset used to calibrate the static model, so this test does not probe slow drifts of the instrument between calibration and observation.
The resulting PSF models and their residuals are shown in \autoref{fig:fm}.

The forward modelled PSFs reproduce the science images well, though worse than when we fit the digital twin directly to the science images, and some of the residuals show correlation with the introduced aberrations.
The discrepancy is due to the Zernike coefficients recovered from the FLOWFS images not exactly matching the true low-order aberrations in the system, despite including the fit of the static offset, and is currently an active part of our research.
Despite this the forward model gives us residuals at $6\times10^{-5}$ at $5\lambda /D$, a factor 5 improvement compared to not subtracting the model, without the science data constraining the time-varying wavefront.

\section{RECOVERING AN INJECTED COMPANION}
\label{sec:companion}

To test whether the forward model preserves companion signal, we add a simulated off-axis PSF to the science images at $\sim 5\ \lambda/D$ with a peak intensity of $10^{-3}$ relative to the unocculted stellar core.
We then run the forward-modelling procedure of \autoref{sec:fm} unchanged on the injected data and subtract the resulting PSF model.

\begin{figure}[ht]
	\centering
	\includegraphics[width=0.45\textwidth]{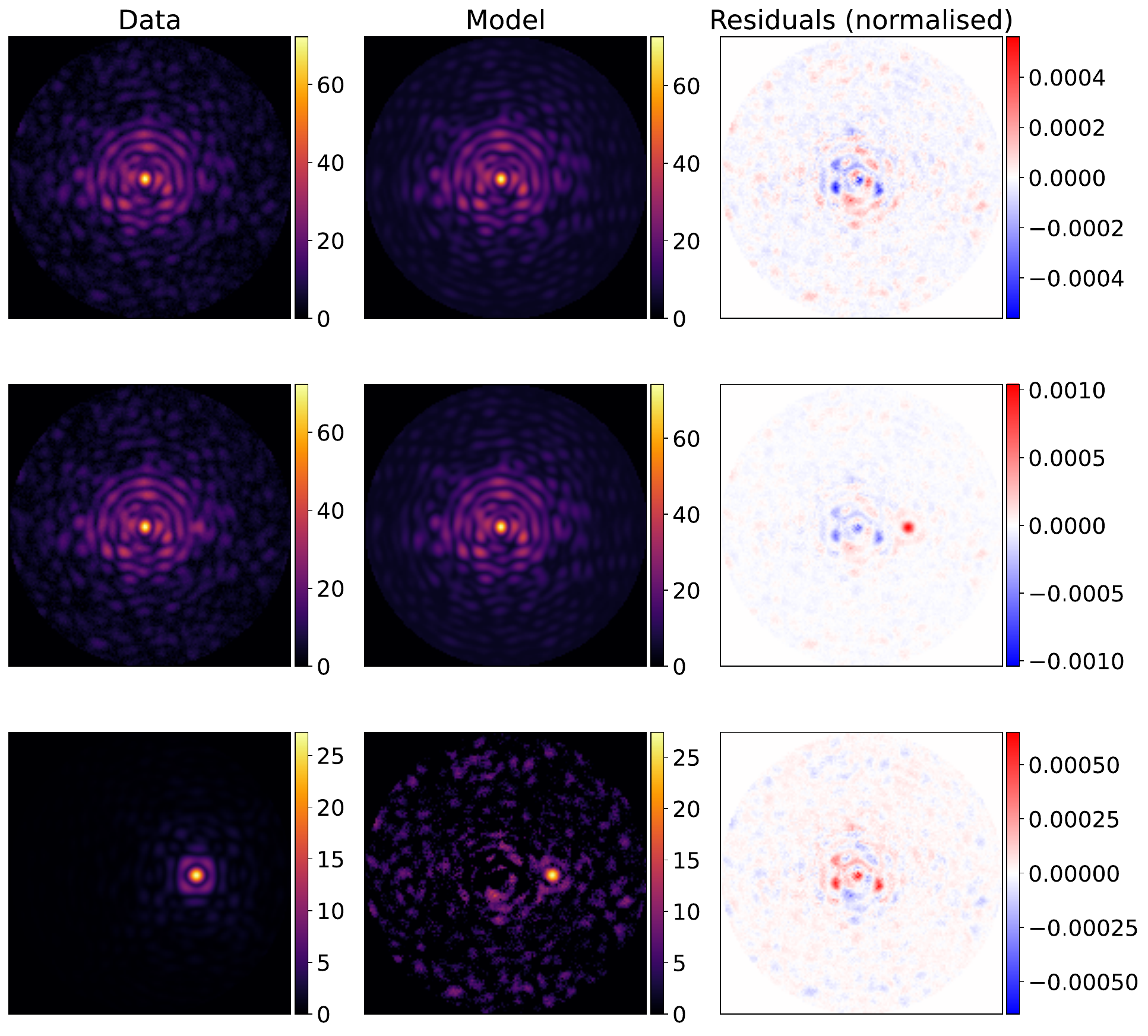} 
	\includegraphics[width=0.40\textwidth]{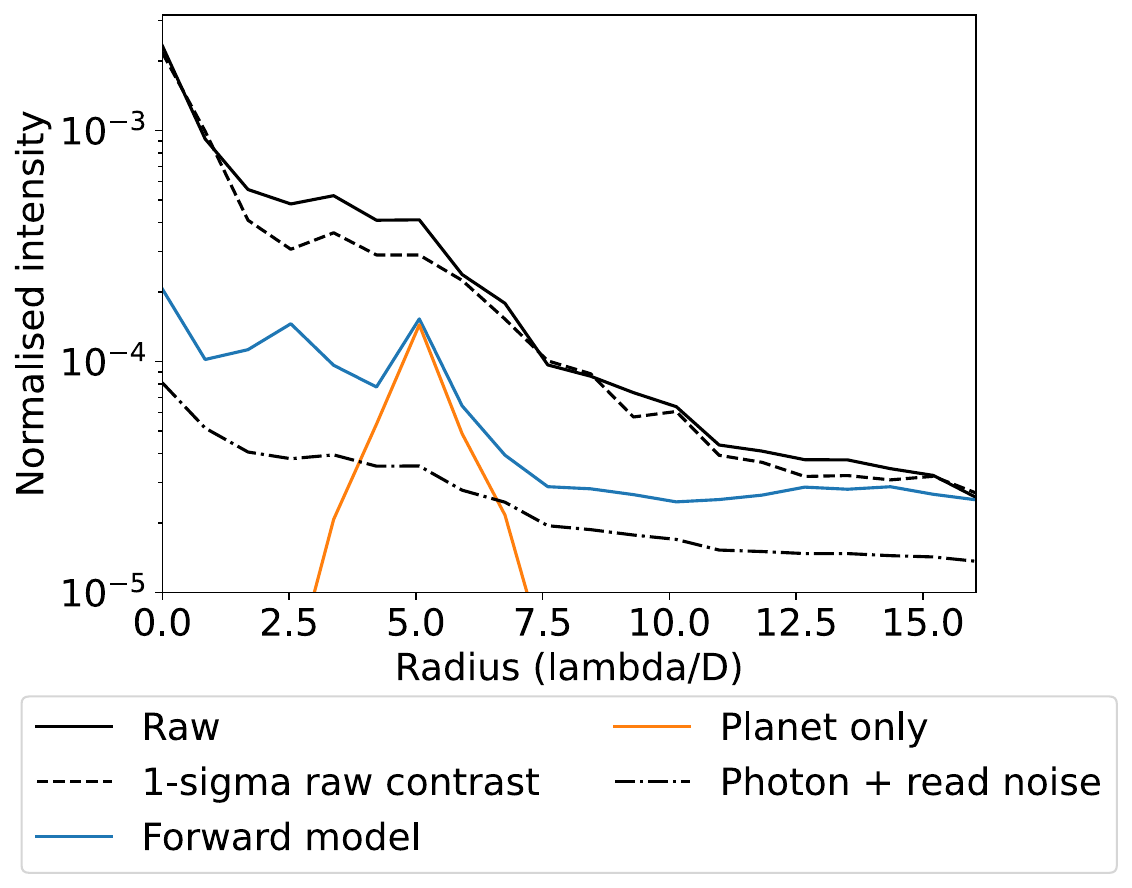}
	\caption[companion]{\label{fig:companion}
		Recovery of an injected companion at $5\ \lambda/D$ with a peak contrast of $10^{-3}$.
		Left, top row: the science image without the companion, the PSF forward modelled from FLOWFS telemetry, and the normalised residual.
		Left, middle row: the same for the image with the companion injected, where the companion now stands out clearly in the residual while the stellar residual is unchanged.
		Left, bottom row: the injected companion on its own (left), the residual in which it is recovered (centre), and the normalised difference between the two (right), which shows no systematic loss of companion flux.
		Right: the standard deviation of the residual per annulus for the injected case, with the companion signal alone plotted for comparison.
		Note that because we plot the standard deviation of the normalised intensity in the annuli, the peak of the companion signal in this right plot is not expected to be at $10^{-3}$, while we do see a $10^{-3}$ signal in the residuals plot on the left.
		The companion is recovered with no measurable self-subtraction, as expected since the stellar model is constrained by the FLOWFS telemetry and has no capability to model an incoherent off-axis point source.
	}
\end{figure}

\autoref{fig:companion} shows the result, with the companion appearing clearly in the residual image. 
When compared to the injected signal we see that we recover all companion signal and do not suffer from oversubtraction.
This is expected from the forward-modelling strategy: as argued in \autoref{sec:strategy}, the model only produces coherent stellar light and its aberrations are inferred from the FLOWFS telemetry, while the remaining free parameters, the per-frame tip/tilt, flux and background and the static low-order phase offset, have no capability to model an incoherent off-axis point source.
Self-subtraction is therefore suppressed by construction.

\section{DISCUSSION AND OUTLOOK}
\label{sec:discussion}

We have shown that a fully physical, differentiable digital twin of MagAO-X can predict the coronagraphic science PSF from FLOWFS telemetry alone, and that subtracting this prediction removes the stellar light without removing companion signal.
In the laboratory, the forward-model residuals reach $6\times10^{-5}$ at $5\ \lambda/D$, a factor of 5 below the raw contrast, and an injected companion is recovered without measurable self-subtraction.
When the same model is fitted directly to the science images it reaches the photon and read noise floor, indicating that the model is capable of representing these data and that the remaining gap in \autoref{fig:fm} is due to how the wavefront estimate transfers between the two branches.

Current research aims to bridge this gap in performance. 
We need a more accurate DM model that uses the measured influence functions rather than assuming that a Zernike command produces a Zernike phase, and a focal plane mask model that includes the reflection off, and transmission through, the glass substrate.
We also need to propagate broadband rather than monochromatic light and model the chromatic effects that follow, in particular in the FLOWFS branch where the bandpass is widest.
A detector model may also be required.
Since the model is mostly physical, the transition to broadband should mainly involve evaluating the same propagation at multiple wavelengths, with chromatic effects modelled outside of the main optical propagation.

Several steps remain before this technique can be demonstrated on-sky.
We have already demonstrated the performance of a FLOWFS digital twin for wavefront sensing and control on sky, running in closed loop at kHz speeds \cite{marsFLOWFS2026}.
What remains is the science-beam model under atmospheric seeing, which will contain higher-order aberrations than the 9 modes that the FLOWFS can sense through its small focal plane mask.
While modelling just the low-order aberrations will already improve the achievable post-processed contrast, we are also developing a digital twin of the pyramid wavefront sensor, in order to provide high-order aberration estimates alongside the low-order ones.
How best to combine two wavefront sensors that run at different frame rates and have overlapping sensitivity ranges is still an open question.
A further practical challenge is the volume of telemetry: a FLOWFS camera reading out $32\times32$ pixel, 16-bit images at 4\,kHz produces roughly 30\,GB of data per hour of observation, which any telemetry-based scheme will have to store, synchronise and reduce.

Finally, none of what we have described is specific to MagAO-X, or to ground-based instruments.
Differentiable digital twins are already being used to calibrate PSFs on space-based instruments \cite{desdoigtsAMIGODataDrivenCalibration2025,fengExoplanetDetectionDifferentiable2025}, and telemetry-based post-processing has been proposed as a route to photon-noise-limited contrast \cite{guyonHighContrastImaging2022}.
The same techniques apply to the Extremely Large Telescopes and to space observatories such as the Habitable Worlds Observatory, where the stability of the instrument should make the digital twin easier to calibrate.

\acknowledgments
The MagAO-X Phase II upgrade program is made possible by the generous support of the Heising-Simons Foundation.
We are very grateful for support from the NSF MRI Award \#1625441 (MagAO-X).
MagAO-X uses the CACAO software package, which is supported by NSF Award \#2410616.
MM, LD and SYH acknowledge support from NWO Award 184.036.004.
SYH acknowledges support from NASA APRA award 80NSSC24K0288.

\bibliography{lib,inprep} 
\bibliographystyle{spiebib} 

\end{document}